# Complementarity between ILC250 and ILC-GigaZ

A. Irles[a], R. Pöschl[a], F. Richard[a] and H. Yamamoto[b].

a Laboratoire de l'Accélérateur Linéaire (LAL), Centre Scientifique d'Orsay, Université Paris-Sud XI, BP 34, Bâtiment 200, F-91898 Orsay CEDEX, France
b Department of Physics, Tohoku University, Sendai 980-8578, Japan

___________________________________________________________________

*Abstract:* In view of the very precise measurements on fermion couplings which will be performed at ILC250 with polarized beams, there is emerging evidence that the LEP1/SLC measurements on these couplings are an order of magnitude too imprecise to match the accuracies reachable at ILC250. This will therefore severely limit the indirect sensitivity to new resonances and require revisiting the possibility to run ILC at the Z pole with polarized electrons. This work was done as a contribution to the **ESU 2018-2020**.

## Introduction

An alternative source of positrons is under consideration by ILC which uses an auxiliary S-band accelerator operating at 3 GeV [1]. It would then be easy to run this machine at the Z pole with only polarized electrons. From a preliminary work performed by H. Yamamoto and collaborators, one can hope to achieve **0.7x10$^{34}$ cm² s$^{-1}$** luminosity at the Z pole, that is two orders of magnitude above LEP1 and comparable to the TESLA project, 0.5x10$^{34}$cm²s$^{-1}$. With e-driven positron source, the damping time limits the collision rate to about 6 Hz. A horizontal emittance smaller by 1/2 with respect to the TDR should be available at Z just like for 250 GeV. The number of bunches can be multiplied by two but the luminosity increase by disruption effect is reduced at Z compared to 250 GeV, so that the luminosity goes like ~ Ecm$^{1.5}$.

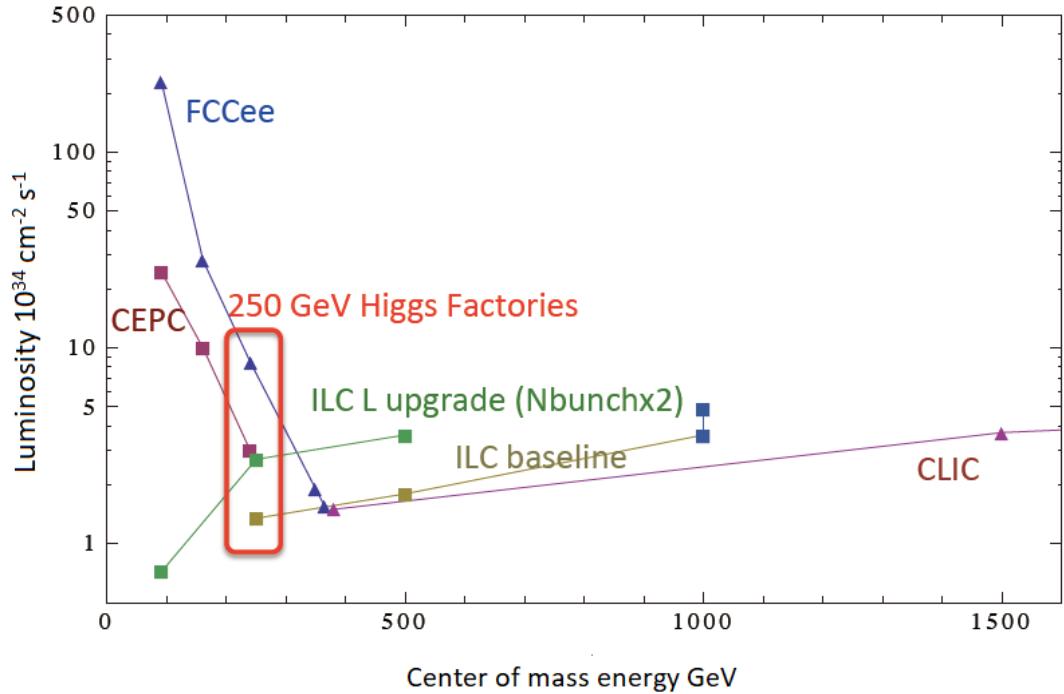

*Figure 1: Comparison between luminosities vs. energy. FCC-ee adding two detectors (dark blue), CEPC (magenta), CLIC (pink) and high-luminosity ILC scenario (green). ILC green is extrapolated to Z as explained in the text*

Figure 1 provides the energy dependence of the ILC luminosity for its base-line and for its high current operation. The base-line assumes spills of 1312 bunches at 5 Hz. To collect 2000 fb-1, the ILC250 assumes a progressive increase to 2624 bunches and 10 Hz [2]. For the operation at the Z pole, one assumes 2624 bunches at 6 Hz.

In this paper, we intend to re-assess the need to go beyond the LEP1/SLC accuracies on Z couplings measurements. We will use as examples some predictions for new heavy Z' resonances based on the Randall Sundrum model. This model offers a well-motivated approach to solve problems which plague the SM, in particular the **hierarchy problem** and the arbitrary appearance of the Higgs field. We will show how ILC250 and ILC-GigaZ, with polarized beams and large currents, can provide the necessary accuracies to test these models well beyond the sensitivity of LHC.

## I. ILC-GigaZ: $\sin^2\theta^\ell_{eff}$ measurement

This part is a reminder of the work done for TESLA in the so-called GigaZ scenario [3].

One can in principle repeat the same arguments with, however, an important difference: to reduce cost and complexity, one assumes that only e- are polarized and therefore the measurement of polarisation comes from the polarimeters and cannot be confirmed by the Blondel method, as when e+ are polarized.

It is however true that ILC will be in a much better situation than SLC (which claimed 0.5% precision on Pe-) since, at 250 GeV, it will be possible to confront the polarimeter results with the **WW measureme**nts. This will no doubt reduce polarimetry uncertainties. To get a comparable precision to TESLA where one uses the Blondel scheme with positrons polarized at 20%, one would need to measure Pe- to 0.05%, an order of magnitude more precise than at SLC. This gain still needs to be demonstrated but our confidence is based on the assumption, previously mentioned, that polarimetry can be calibrated at 250 GeV.

As for Tesla, one measures:

$$ALR=NL-NR/(NL+NR)Pe-=2v_ea_e/(a_e^2+v_e^2)=A_e$$

$a_e$=-0.5 and $v_e$=-0.5+2$\sin^2\theta^\ell_{eff}$ being the vector and axial vector couplings of the electron to the Z.

All hadronic decay modes are used and therefore GigaZ gives $\delta ALR=3\times10^{-5}$, which requires $\delta Pe-/Pe-\sim2\times10^{-4}$ to match the statistical accuracy. As discussed below, the monitoring of the beam energy spread gives a worse accuracy, $\delta ALR=10^{-4}$. To reach this accuracy, the TESLA TDR takes an equivalent polarisation error of 0.05% which is achieved by using positrons with 20% polarisation [4].

In what follows one will assume for ILC-GigaZ :

$$\delta Pe-/Pe-=0.05\%$$

which is a factor 10 better than achieved at SLC but, again, without the resources of calibrating the polarimeters using the WW channel.

The corresponding accuracy is $\delta\sin^2\theta^\ell_{eff}$ **=1.3x10$^{-5}$**, to be compared to the present world precision, 1.3x10$^{-4}$, and two times worse than 6x10$^{-6}$ claimed by FCCee with ~300 times more luminosity. Note in passing that the **compensating gain due to polarisation** was already observed with SLC that could overcome LEP1 results with ~30 times less integrated luminosity.

FCCee uses only the cleanest mode ee->μμ to measure **AFBμ,** while ILC with polarisation can include all hadronic modes to measure the polarisation asymmetry **ALR**. Using AFBμ, FCCee has access to the combination **AeAμ** and not to Ae individually, as will be the case with ILC. It has therefore to assume **lepton universality, Ae=Aμ, to extract Ae and $\sin^2\theta^\ell_{eff}$.** This assumption is by no means trivial and we will come back to this in the last section.

## II. Systematics

Precision measurements at GigaZ are usually not limited by statistical errors but by various instrumental and theoretical effects.

### II.1 Calibration in energy

A constraint to be remembered is the beam energy spread, which for certain EW measurements will be essential. Recall, as rightly reminded by the FCCee CDR, that the asymmetry measurements are very sensitive to the beam energy error through the influence of **photon-Z interference effect**.

For TESLA, it was assumed that there would be energy monitoring using the beam magnetic spectrometers. Quantitatively one should require a control on the energy which avoids degrading the accuracy on ALR. One predicts:

$$dALR/d\sqrt{s}=2\times 10^{-5}/MeV$$

where this measurement uses all hadronic final states, u, d, s, c, b (see Appendix). Note in passing that this dependence is steeper if one selects lepton final states. Note also that this dependence will be an order of magnitude reduced for quantities like Rb or Rℓ.

Reaching a $10^{-4}$ accuracy on ALR is achievable with a spectrometer resolution of ~1 MeV as claimed by TESLA. The spectrometer can be calibrated using an energy scan around the well-known Z resonance. Beamstrahlung should also be kept under control since it produces a $9\times 10^{-4}$ shift on ALR, again assumed to be feasible in the TESLA TDR [4]. The error quoted by TESLA is therefore achievable: **δALR=10⁻⁴**.

For what concerns the quantity **ALRFB** which, as recalled in the Appendix, is used to extract Ab, Aµ and Aτ, the energy dependence is similar. These results are summarized in the following table:

| Type of fermions | µ,τ | b | c | All quarks |
|---|---|---|---|---|
| **dALR/d√s 10⁻⁵/MeV** | 6 | 1.92 | 2.11 | 1.99 |
| **dALRFR/d√s 10⁻⁵/MeV** | 0.9 | -0.34 | | |

## II.2 Polarisation measurement

At SLC, the positrons were produced from an energetic polarized electron beam (J. Seeman, private communication), which may have induced a remnant polarisation of the positrons. This effect was however unproven. At the ILC-GigaZ, the positron beam can be conventionally produced from an unpolarized 3 GeV electron beam. Therefore it is justified to suppose that the polarisation of the positrons is exactly 0 since no mechanism can be invoked to generate positron polarisation.

## II.3 Theoretical errors

In [5], a review of theoretical errors has been presented and is recalled in the table below. For Rb, discussed in the next section, the expected error will be $1.5\times 10^{-4}$, while the experimental error for ILC is of the same order. One also observes that the theoretical uncertainty on $\sin^2\theta^\ell_{eff}$ is **3 times the expected accuracy at ILC-GigaZ, 1.3x10⁻⁵**. For Ab, the theoretical uncertainty $\delta Ab=3.3\times 10^{-5}$ is smaller than our expected accuracy $5\times 10^{-4}$.

| | $\delta\Gamma_Z$ [MeV] | $\delta R_\ell$ [$10^{-4}$] | $\delta R_b$ [$10^{-5}$] | $\delta\sin^2\theta^\ell_{eff}$ [$10^{-6}$] | $\delta\sin^2\theta^b_{eff}$ [$10^{-5}$] |
|---|---|---|---|---|---|
| Present EWPO errors | | | | | |
| EXP1 [11] | 2.3 | 250 | 66 | 160 | 1600 |
| TH1 [9, 38, 39] | 0.5 | 50 | 15 | 45 | 5 |
| FCC-ee-Z EWPO error estimations | | | | | |
| EXP2 [40] & Tab. 2 | 0.1 | 10 | 2÷6 | 6 | 70 |

This table recalls the accuracies expected at FCCee [5]. Extrapolations to better TH errors are expected in a foreseeable future.

# III. Asymmetry measurements at ILC-GigaZ

TESLA assumes δRb=0.21653±0.00014. The gain in accuracy at FCCee very much depends on assumptions on **systematics** and can be as low as a **factor two** with respect to ILC.

Recall that for **Rb,** the luminosity error cancels on this ratio, while the b-tagging efficiency evaluation follows methods developed at LEP1 (ratio of double to single tag) but with improved b-tag efficiencies given the smaller radius of the ILD beam pipe, analogue to SLD.

The measure of Ab at SLC uses the following combination of measurements (see Appendix):

$$ALRFB = \frac{1}{Pe} \frac{(\sigma F - \sigma B)L - (\sigma F - \sigma B)R}{(\sigma F + \sigma B)L + (\sigma F + \sigma B)R} = \frac{3}{4} Af$$

This quantity directly provides Ab when selecting the bb channel, noting that for this channel one can define forward and backward events by selecting charged B mesons or charged kaons, as currently performed with ILD [6].

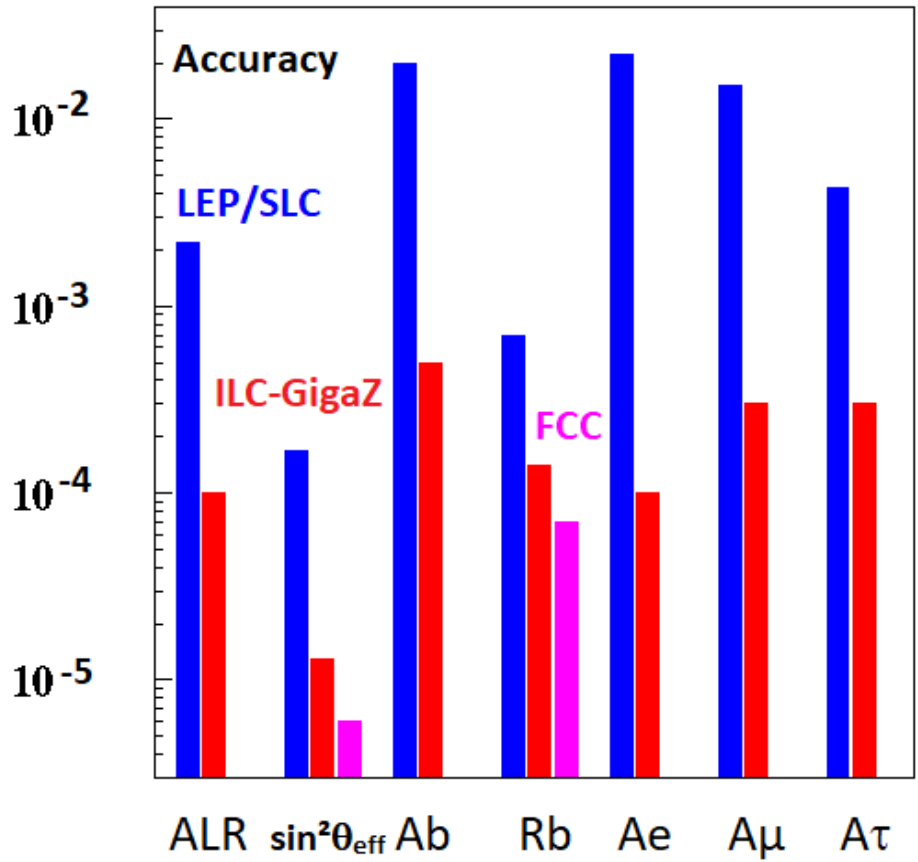

*Figure 2 : Estimated measurement errors at ILC-GigaZ compared to LEP1/SLC and FCCee*

The accuracy on Ab can be understood easily by extrapolating from the result of SLD which gives an error on Ab of 2% with 0.5x10$^6$ hadronic events. ILC-GigaZ will collect 0.7x10$^9$ hadronic Z decays, hence an error of 0.05%. With better tracking and kaon identification in ILD, one expects a statistical error ~ 0.015%, meaning that the polarisation error will become dominant.

As for ALR, one has checked how this various methods are affected by energy spread.

Figure 2 summarizes what can be expected from ILC-GigaZ, showing an impressive progress with respect to LEP1/SLC.

## IV.  BSM physics from ILC250 with ILC-GigaZ measurements

The reactions **ee->bb** and **ee->cc** are been studied at ILC250 [7]. Present results are based on a new study, under preparation, using 2000 fb$^{-1}$ at ILC250 and increased efficiency. Polarized beams allow to separate the 4 different chirality combinations **LeLb, LeRb, ReRb, ReLb** (and similar quantities for c quarks). As will be shown on examples, these four modes can be differently influenced by BSM physics and their separation allows to identify the underlying mechanism and extract MZ'. These four amplitudes can be decomposed as follows:

$$\text{LeLb} = QeQb + \frac{LeZLbZ}{s^2wc^2w}BWZ + \sum_{Z'}\frac{LeZ'LbZ'}{s^2wc^2w}BWZ'$$

ILC250     SM     GigaZ                          New resonances

where the charges are Qe=-1, Qb=-1/3, the SM Z couplings are LeZ=Ie3-Qes²w, LbZ=Ib3-Qbs²w and BWZ=s/(s-MZ²) is the Breit Wigner of the Z. One can write similar formulae for the three other combinations LeRb, ReRb and ReLb.

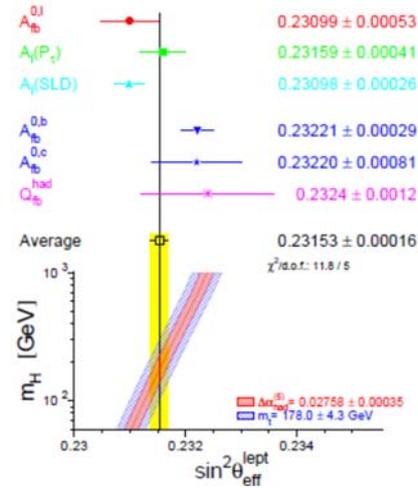

Figure 3 Summary of LEP1 and SLC results for the measurement of $\sin^2\theta^{lep}_{eff}$.

LeZ' and LbZ' are provided by BSM models, as illustrated by figure 4 given below.

While the first term, the photon exchange, should remain unaltered by virtue of the U(1)em symmetry, the second term can be affected by Z-Z' mixing or b-b' mixing. As is well known, this channel has provided the most significant deviations at LEP1 and in [8] Z-Z' mixing has been invoked to explain the deviation on AFBb.

If one considers, figure 3, the two most precise measurements at the Z pole providing sin²θ$^\ell$$_{eff}$, ALR from SLC and AFBb from LEP1, one observes that they differ by 3.5 s.d. This alone shows the interest to re-measure these two quantities with much higher accuracy. Moreover, it would be far preferable to measure Ab directly, as this quantity is unaffected by the Ae behaviour, in contrast to AFBb measured at circular colliders w/o beam polarisation.

Generally speaking, the third term of the equation above contains an ensemble of vectorial heavy resonances which, in the case of Randall Sundrum (RS) models are Kaluza-Klein recurrences of the photon, the Z and new resonances of the type Z', due to additional symmetry groups. The first and third term are negligible on the Z resonance given that there is almost no interference with the Z imaginary amplitude, while this interference can become manifest at ILC250 where all three terms are real. **To unambiguously isolate this new contribution, it is therefore essential to measure accurately the second term at the Z resonance.** In the absence of a run at the **Z pole**, one would be left with the unsatisfactory LEP/SLC measurements, with insufficient accuracy **to match in precision the measurements performed at ILC250** and would therefore limit the sensitivity of ILC to new physics, as explained by a numerical example in the next section.



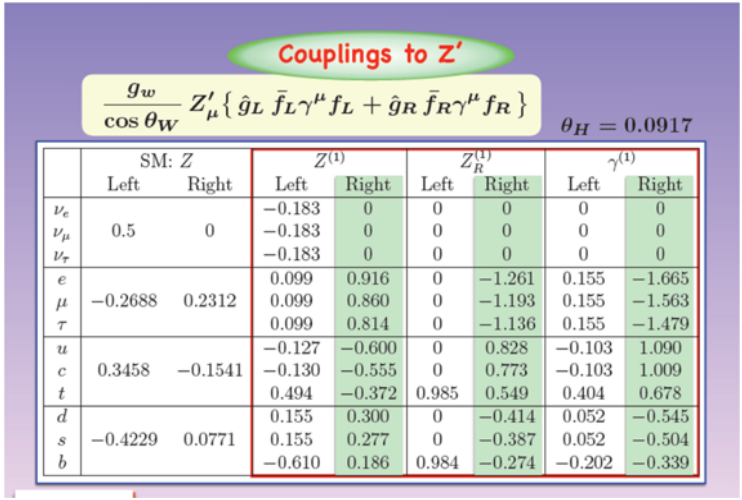

One would still be able to witness the presence of BSM physics but unable to isolate the contribution from new resonances (in red in above equation). **The essential mission of the ILC is however to discover new physics and provide information on the masses of new resonances for future colliders.**

To get a feeling of what can be expected from BSM, one can use above table from [9], a BSM model with three heavy resonances with masses from 7 to 9 TeV.

*Figure 4 : Predicted couplings in [9] for the 3 heavy bosons. The first column corresponds to the SM couplings*

From figure 4, one observes that both **bR** and **Lb** couplings to the heavier resonances can be up to four times the SM coupling while the **eR** coupling reaches 7 times the SM coupling. This explains how ILC250 can reach such a sensitivity on the **ReRb** and **ReLb** combinations as shown by figure 5. The **cR** coupling is even larger and with similar reconstruction efficiencies with the ILD set up, one expects a larger effect as shown in figure 5.

Moreover these effects will also be marked for the **right handed leptonic** couplings, including the ee->ee reaction which requires a more subtle analysis [11].

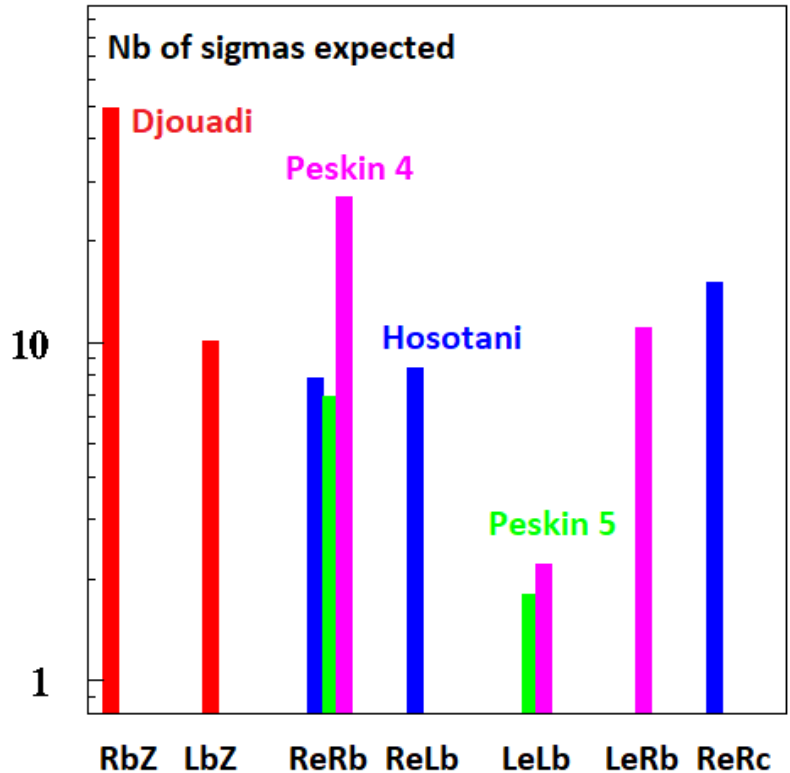

*Figure 5: Predicted number of standard deviations for ILC-GigaZ + ILC250 for the 3 models described in the text.*



# V. Results

Using three explicit RS models based on extra dimensions, figure 5 shows the expected number of standard deviations from a combination of an ILC-GigaZ measurement giving RbZ, LbZ, RcZ, LcZ from and the 4 helicity combinations measured at 250 GeV. The Djouadi model [5] was designed to reproduce the LEP1 effect and corresponds to a Z' with a ~3 TeV mass which mixes with Z. This Z' does not couple to e+e-, and the KK excitations of the photon and the Z are very weakly coupled. The model of Hosotani et al. [9], already discussed in the previous section, assumes KK resonances at 7-9 TeV and no mixing at the Z resonance. Peskin and Yoon provide two hypotheses referred as Peskin 4 (bL in 5 and bR in 4 of SO(5)) and Peskin 5 (bL and bR in 5 of SO(5)).

These results suggest that ILC250+GigaZ can discover RS signals and, more importantly, pick up the right underlying model. Then these precision measurements would allow to predict indirectly the mass of the new resonance. How far in mass can ILC reach is discussed in the next section.

## V.1 Sensitivity to very high Z' masses

At a modest energy of 250 GeV, ILC has a reach which extends well above LHC direct searches. For what concerns the Hosotani model, the large deviation which would be seen at ILC for a 8 TeV resonance indicates that ILC250 can extend it sensitivity even far beyond this mass. Since these resonances are very wide, LHC, is already able to exclude them up to ~9 TeV [9]. In the absence of a signal at ILC250, one would conclude that such a **Z' is heavier than 20 TeV at 95% C.L.** for the b channel, reaching **25 TeV** for charm. This limit could reach ~**30 TeV** with ILC500, taking into account that the gain in luminosity, less than 2, does not compensate for the drop of the cross section ~5.

Furthermore, as discussed in [11], leptonic channels, with beam polarisation and with ~100% reconstruction efficiency and negligible background can also reach **mass limits above 20 TeV** at ILC250.

These figures illustrate how ILC could pave the way to the design of a pp collider which would be able to observe directly such heavy resonances.

## V.2 The need for ILC-GigaZ

Recall again **that effects from Z' propagators can only be established if we re-measure the lepton and b couplings at ILC-GigaZ.** To see this quantitatively, let's compare the measurement error on the amplitude **ReRb** which has, according to above plot, the **highest occurrence of deviations**, to the uncertainty due to the b couplings measurement at LEP1. At ILC250 one has:

$$ReRb = Q_e Q_b + R_e Z R_b Z B W Z (s^2 w / c^2 w) = 1/3 + 0.115 \text{ with } BWZ = s/(s - M^2_Z)$$

where the second term, ~0.115, is the Z contribution. To match the expected measurement accuracy with 2000 fb$^{-1}$ collected at 250 GeV, δ**ReRb~0.0007**, one needs a relative accuracy on **Rbz** of 0.0005/0.115=0.7%, while the present LEP1 measurement gives ~10%. In the absence of a GigaZ baseline, one would be able to observe a **significant deviation from the SM** but be totally **unable to decide** if it comes from an alteration of SM Z couplings or from the presence of Z' propagators. This would result in a **major waste of information**.



At ILC-GigaZ, one expects: $\delta RbZ/RbZ=0.55\%$ (and $\delta LbZ/LbZ=0.04\%$) which **perfectly matches the accuracy achievable with ILC250.**

This discussion illustrates the **specific role of b quark** measurements with respect to **top quark** measurements: they allow a perfect separation between Z coupling anomalies and propagator contribution. To achieve the same separation for top quarks one needs to run at two energies above the top threshold.

## V.3 Summary

These examples clearly show:

- The importance of running at the Z pole to measure the Zbb couplings, eventually detecting an anomaly which needs to be taken into account for the measurement at 250 GeV
- The sensitivity of ILC to detect **the presence of Z'** propagators at 250 GeV, provided the Zbb couplings are re-measured with ILC-GigaZ.
- The importance of the measurements with polarized beams to disentangle the four amplitudes **LeLb, LeRb, ReRb, ReLb,** allowing to discriminate between the various models
- Charm and lepton couplings are also relevant in the Hosotani scheme.

# VI. The lepton universality and BSM physics

Recalling the various anomalies observed at B factories where, e.g. b->s$\mu\mu$ differs from b->see, it will be important to verify **lepton universality** at the Z pole and at ILC250.

At the Z pole, it will be straightforward to measure R$\mu$ and A$\mu$ using the same observables as for b quarks. The rate is ~5 times smaller but the efficiency is close to 100% for both measurements which gives an almost complete compensation for the efficiency for b quarks. One has **$\delta R\mu/R\mu=0.017\%$** and **$\delta A\mu=0.03\%$**.

At FCCee, one can measure the ratio R$\tau$/R$\mu$ with a relative error of $5\times10^{-5}$ which allows a stringent test of universality for the combination **$g^2L\ell+g^2R\ell$,** while ILC-GigaZ allows to measure **$(g^2L\ell-g^2R\ell)/(g^2L\ell+g^2R\ell)$** which may show different deviations.

At ILC250, one will be able to measure separately **ReR$\mu$, ReL$\mu$, LeL$\mu$ and LeR$\mu$** and the same is true for $\tau$ leptons which, together with the b coupling measurements, will also allow to identify the origin of BSM contributions (see e.g. [9]). **For ee->$\mu\mu$ and ee->$\tau\tau$ one expects ~30 s.d**. on ReR$\mu$ and ReR$\tau$ in this model. The uncertainty due to Z couplings is negligible given the accuracies expected from ILC-GigaZ but this would not be the case if one were left with LEP/SLC accuracies ten times worse.

## Conclusion

From this brief survey, one can conclude that ILC-GigaZ is a necessity which should not be disregarded since it offers a powerful access to **EW precision measurements,** comparable to FCCee.



For the emblematic measurement of **sin²θ$^ℓ_{eff}$**, it allows to improve the present accuracy by a factor 10, **beyond present theoretical uncertainties.**

This achievement requires either polarized positrons, at the expense of a more complicated scheme, or, as for SLC, an electron beam with a polarisation measured **10 times better than at SLC**. While very challenging, the latter goal does not seem out of reach given that polarimetry can be cross-checked at ILC250 where one can use the presence of the WW channel and, eventually, the Blondel method if there are polarized positrons at 250 GeV.

As shown in a concrete example, these measurements are complementary to those performed by ILC250. This example also demonstrates that **beam polarisation** plays an essential role in distinguishing between various RS models.

In the absence of a run at the Z pole, one would be left with the LEP/SLC measurements accuracies which would not match in precision the measurements performed at ILC250 which would therefore decrease the sensitivity of ILC to new physics and, more importantly, limit the capability of ILC250 to predict the masses of the new resonances.

> ➔ **Running ILC-GigaZ is therefore required to benefit from the high accuracies provided by ILC250 on fermion coupling**

Not to be forgotten is the issue of **HO theoretical corrections**, in particular EW corrections which for the top channel have revealed very significant effects [13]. Given that experimental errors will reach the 0.1% level, it is essential to launch an important effort on this topic.

Note in passing that ILC-GigaZ with polarisation can provide a rigorous and unique test of **lepton universality** by measuring separately **Ae, Aµ and Aτ**. This test could be crucial given some departures of lepton universality observed in B factories.

This ensemble of results would constitute a major step forward in our comprehension of **fermion properties** with a major potential for **BSM discoveries**.

**Acknowledgements.** *This work has received encouragements from several ILC colleagues whom we thank gratefully. For machine aspects, H. Yamamoto would like to thank Kaoru Yokoya, Shinichi Michizono, and Benno List for useful contributions to the luminosity issue. We acknowledge the precious help from the ILD physics and software working groups, in particular from F. Gaede and A. Miyamoto for efficiently providing high quality simulated data. Finally we thank the organizers of the Linear Collider Community Meeting 8-9 April 2019, in Lausanne, to have given us an opportunity to present these results. Finally F.R. is grateful to A. Blondel for pointing out a mistake in the calculations of Aµ and Aτ at ILC-GigaZ.*

# APPENDIX

## Asymmetries

Recall here the methods used by SLD to extract Ab, see [12].

The Born level differential cross-section, with a polarized electron beam, is given by:

$$\frac{d\sigma b}{d\cos\theta} = \frac{3}{8}\sigma b[(1 - P_e A_e)(1 + \cos^2\theta) + 2(A_e - P_e)A_b \cos\theta]$$

Where the electron beam polarisation Pe is positive for right-handed beam. A forward-backward-left-right asymmetry can be formed as:

$$A_{LRFB} = \frac{1}{P_e}\frac{(\sigma F - \sigma B)L - (\sigma F - \sigma B)R}{(\sigma F + \sigma B)L + (\sigma F + \sigma B)R} = \frac{3}{4}A_f$$

This measurement can be applied to all fermionic final states where one can measure the fermion charges: µ, τ, b and c. In particular it gives Ab.

It is directly dependent on the electron polarisation.

The dominant systematical error comes from Pe (luminosity and efficiency go away on this ratio): δAb/Ab=δPe/Pe. Taking 0.05% on Pe, on gets δAb=0.0005.

AFBLR=0.6Ab~0.85.

    var(AFBLR)=var(F-B)L+var(F-B)R))/(NL+NR)²= NL(1-AFBL²)+NR(1-AFBR²)/(NL+NR)²

AFB=2(Ae-Pe)Ab/(8/3)(1-PAe) gives AFBR=-0.674 and AFBL=0.592 for P=0.8. Assuming a 30% efficiency, this gives δAFBLR/AFBLR=1.37 10$^{-4}$ which is almost negligible with respect to the systematical error, hence:



$$\delta A_b = 0.0005$$

$A\mu$ and $A\tau$ are ~0.151 and the efficiency is ~100%. $A_{FBLR}=0.6A\mu$. The asymmetries are almost negligible giving for the statistical term $var(A_{FBLR})=1/(N_L+N_R)=3 \cdot 10^{-8}$ hence $\delta A\mu=3 \cdot 10^{-4}$. The error due to Pe, gives $\delta A\mu=\delta A\tau=0.75 \cdot 10^{-4}$, almost negligible, hence:

$$\delta A\mu = \delta A\tau = 0.0003$$

Energy smearing is negligible at this level of accuracy.

## Energy dependence

For ALR and ALRFB one needs to worry about the energy dependence induced by the photon-Z interference. Due to **beamstrahlung** and beam energy resolution, this effect cannot be eliminated by sitting on the resonance. Detailed computations will not be reproduced here. To evaluate analytically the differential dependence of these quantities, the principle is rather simple. One can easily show that:

$$dA_{LR}=[LL'^2+LR'^2-RR'^2-RL'^2-2Q_f(LL'+LR'-RR'-RL')(dm^2/m^2)]/[(LL'^2+LR'^2+RR'^2+RL'^2-2Q_f(LL'+LR'+RR'+RL')(dm^2/m^2)]$$

Where, for instance, $LL'=(-0.5+s^2w)(I_{3f}-Q_fs^2w)/s^2wc^2w$ $I_{3f}$ and $Q_f$ are the isospin and the electric charge of the final fermions. Adding all quark flavours, one finds:

$$dA_{LR}/d\sqrt{s} = 1.99 \times 10^{-5}/MeV$$

in very good agreement with the TESLA TDR. It also appears that in the expression of ALR=(L-R)/(L+R), the main energy dependence comes from the numerator while the denominator shows a variation more than ten times slower. This means that the Rb measurement will show a negligible effect.

One could measure ALR using lepton final states but in this case the derivative is larger:

$$dA_{LR}\mu/d\sqrt{s} = 6.2 \times 10^{-5}/MeV$$

This method also allows to compute the energy variation of ALRRFB:

$$dA_{LRFB}b/d\sqrt{s} = -0.34 \times 10^{-5}/MeV \quad \text{and} \quad dA_{LRFB}\mu/d\sqrt{s} = 0.9 \times 10^{-5}/MeV$$

showing a similar dependence as ALR.